\documentclass[conference]{IEEEtran}

\usepackage[paperwidth=8.5in,paperheight=11in,left=1.0in, right=1.0in, top=1.5in, bottom=1.5in]{geometry}

\usepackage{listings}

\usepackage{verbatim}

\usepackage{multicol}

\usepackage{enumerate}
\usepackage{graphicx} 
\usepackage{url}
\usepackage{amssymb}
\usepackage{color}
\usepackage{ifthen}
\newboolean{showcomments}
\setboolean{showcomments}{true}
\ifthenelse{\boolean{showcomments}}
 { \newcommand{\mynote}[2]{
      \fbox{\bfseries\sffamily\scriptsize#1}
        {\small$\blacktriangleright$\textsf{\textcolor{red}{{\em #2}\bf }}$\blacktriangleleft$}}}
        { \newcommand{\mynote}[2]{}}

\usepackage{hyperref}
\usepackage{tabularx}
\usepackage{xcolor}
\definecolor{delim}{RGB}{20,105,176}
\colorlet{punct}{red!60!black}
\colorlet{numb}{magenta!60!black}
\usepackage{listings}
\lstdefinelanguage{json}{
    basicstyle=\scriptsize\ttfamily,
    numberstyle=\scriptsize,
    showstringspaces=false,
    breaklines=true,
    literate=
     *{0}{{{\color{numb}0}}}{1}
      {1}{{{\color{numb}1}}}{1}
      {2}{{{\color{numb}2}}}{1}
      {3}{{{\color{numb}3}}}{1}
      {4}{{{\color{numb}4}}}{1}
      {5}{{{\color{numb}5}}}{1}
      {6}{{{\color{numb}6}}}{1}
      {7}{{{\color{numb}7}}}{1}
      {8}{{{\color{numb}8}}}{1}
      {9}{{{\color{numb}9}}}{1}
      {:}{{{\color{punct}{:}}}}{1}
      {,}{{{\color{punct}{,}}}}{1}
      {\{}{{{\color{delim}{\{}}}}{1}
      {\}}{{{\color{delim}{\}}}}}{1}
      {[}{{{\color{delim}{[}}}}{1}
      {]}{{{\color{delim}{]}}}}{1},
}
\begin{document}

\title{Model-Driven Analytics: Connecting Data, Domain Knowledge, and Learning}
	
\author{
    \IEEEauthorblockN{Thomas Hartmann, Assaad Moawad, Francois Fouquet,\\ Gregory Nain, Jacques Klein, and Yves Le Traon}
    \IEEEauthorblockA  {
Interdisciplinary Centre for Security, Reliability and Trust (SnT),\\
University of Luxembourg\\
Luxembourg\\
firstName.lastName@uni.lu
    }
    \and
    \IEEEauthorblockN{ Jean-Marc J\'ez\'equel}
    \IEEEauthorblockA  {
        IRISA,\\
University of Rennes I,\\
France\\
firstName.lastName@irisa.fr
    }
}







 
\maketitle	
	
\begin{abstract}
Gaining profound insights from collected data of today's application domains like IoT, cyber-physical systems, health care, or the financial sector is business-critical and can create the next multi-billion dollar market.
However, analyzing these data and turning it into valuable insights is a huge challenge.
This is often not alone due to the large volume of data but due to an incredibly high domain complexity, which makes it necessary to combine various extrapolation and prediction methods to understand the collected data.
Model-driven analytics is a refinement process of raw data driven by a model reflecting deep domain understanding, connecting data, domain knowledge, and learning.

\end{abstract}


\section{If Data is the New Oil, how do we refine it?}
Today, data is generated in very large scale by a wide range of sources, such as sensors, embedded devices, social medias, and audio/video.
Advances in storage technologies and their continuously falling prices allow to collect and store huge amounts of data for a long time, creating entirely new markets aiming at valorizing this data.
Recent studies, for example from McKinsey~\cite{mckinsey}, emphasize the tremendous importance of this relatively new field by calling it the \textit{``next frontier for competition''}. 
Others even compare the value of data for modern businesses with the value of oil, referring to data as \textit{``the new oil''}~\cite{forbes}.  
However, as it is the case for crude oil, data in its raw form is not very useful.
To transform crude oil into value, a long valorization chain, composed of heterogeneous transformation steps, needs to be applied before oil becomes the essential energy source we all so heavily rely on.
Similarly to oil, to turn data into the multi-billion dollar business that some analysts predict it will become~\cite{mckinsey}, we need to process and refine data before we can get valuable insights out of it. 
This process of turning raw data into valuable insights is referred to as \textit{data analytics}.
If we stay with the oil analogy, data analytics can be seen as the refinement process of data. 
It has the potential to help us to better understand our businesses, environment, physical phenomenas, bodies, health, and nearly every other aspect of our lives.
However, turning collected data into competitive advantages remains a big challenge. 
Besides the scale of collected data, it is also the ever increasing complexity of today's application domains like IoT, cyber-physical systems, health care, or the financial sector which makes gaining profound insights form collected data so challenging.
Various methods such as statistical metrics, machine learning (ML), and extrapolation models to predict physical phenomena have to be combined with knowledge of \textbf{domain experts} to create added value out of raw data.

\subsection{The Big Data Pathology}
\begin{figure*}	
	\centering			
	\includegraphics[scale=0.4]{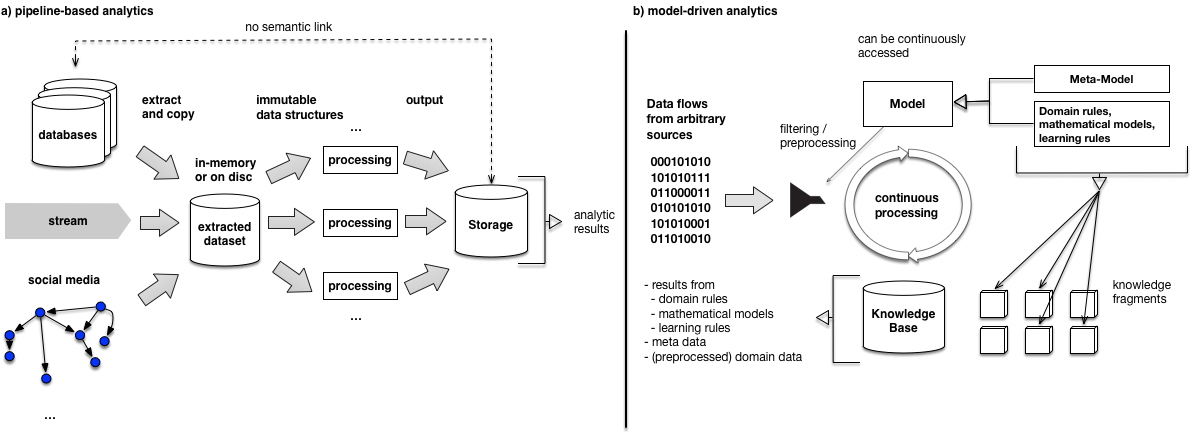}
	\caption{Schematic working principle of a) typical pipeline-based data analytic processes compared to b) model-driven analytics}
	\label{fig:todays_analytics}
\end{figure*}

Most of todays data analytic techniques are processing data in a pipeline-based way: they first extract the data to be analyzed from different sources, {\em e.g.,} databases, social medias, or stream emitters, copy them into some form of usually immutable data structures, stepwise process it, and then produce an output.     
By parallelising the processing steps, {\em e.g.,} based on the map-reduce programming model~\cite{Dean:2008:MSD:1327452.1327492}, these techniques are able to mine huge amounts of data in comparatively little time and can find all kind of useful correlations.
This process is depicted in Figure~\ref{fig:todays_analytics} a). 
\textbf{While this is very suitable for tasks like sorting vast amounts of data, analyzing huge log files, or mining social medias for trends (even in near real-time) it is less suitable for analytics of domains with complicated relationships between data where several different analytic techniques and models need to be combined with domain knowledge and ML to refine raw data into deep understanding}~\cite{crankshaw2014missing}. 
For such analytics a pipeline-based approach has severe drawbacks and easily leads to an inefficient ``blindly store everything and analyze it later'' approach, which is referred to as the ``big data pathology'' or ``big data trap''~\cite{jacobs2009pathologies}.

\subsection{Exemplifying Application Domains}
\label{sec:bigdata_pathology:examples}
As a first example lets consider infrastructure monitoring systems.
We are working in this context on a smart grid monitoring system together with Creos Luxembourg, the main electricity grid operator in the country.
The goal of this work is to continuously monitor the electricity grid through measured and stored data to detect but also predict possible failures, consumption peaks, attacks, and other potential problems in the grid.
For instance, to avoid potential peaks, various statistical forecasts (based for example on learned consumption habits and weather conditions) have to be combined with the underlying electrical laws to anticipate the load in cables and possible side-effects.
Considering the high volatility of live measurements in this domain, it is nearly infeasible or at least extremely expensive to simply restart the whole pipeline process for every changed value to produce a new output. 
However, it is difficult to anticipate what exact impacts a new value could have, {\em e.g.,} consumption profiles, electrical loads, prediction models and so forth may have to be updated.
Moreover, simply storing every measured value without knowing if the value is useful for later analysis further increases computational costs.

Self-driving cars, prominently promoted, among others, by Google, are another example of a complex analytics domain. 
During every ride, these cars need to make numerous decisions based on collected data, an understanding of the current situation, and predictions on future outcomes.
For every decision they have to anticipate what might happen in terms of traffic, pedestrians, road conditions and so forth and need to evaluate the effects of possible decisions before actually deciding for one. 
This again needs to combine data, different analytic tools, such as scientific models defining physical rules for speed, acceleration, braking distances, etc., domain knowledge like the meaning of road signs or crosswalks, and ML to be able to learn from previous situations. 
The questions is, which values have impacts in which circumstances and which parts of the analytics produces the output knowledge how the car has to behave in this situation?

Another domain where complex data analytics is becoming increasingly important is the financial industry.
Live simulations of trading strategies are used to predict their chances of success and the expected profit growth. 
As in the previous examples, this requires the combination of different analytic tools, domain knowledge, ML, and data, {\em e.g.,} real-time stock prices. 
For instance, scientific models can represent financial theories, domain knowledge can describe complex risk models, and ML can help to forecast stock market trends. 
However, it is difficult to anticipate what impacts a change of a risk model or stock value has and which parts of the simulation are impacted and need to be executed again. 

\section{Drawbacks of Todays Pipeline-based Analytics}
\label{sec:drawbacks}
We identify four main drawbacks which make pipeline-based analytics for complex application domains less suitable.

\textit{1)} For every little change in the source data the whole pipeline process of extracting, copying, processing, and producing an output has to be redone.
However, the produced output, {\em i.e.,} the gained knowledge, in this programming model is hardly used to feedback the source data itself. 
This is because there is no semantic definition of the output and how the output could be reused together with new data to create new knowledge.
Neither is there an explicit definition of what parts of the input data led to which output. 
The transformation steps just encode how input data is syntactically transformed into output data.
Therefore, the output has to be completely recomputed from scratch by restarting the pipeline process without reusing previous results. 
Again, this programming model is very suitable for tasks like transforming an input dataset into an output dataset, {\em e.g.,} sorting, or finding distinct values, but is less suitable for domains with many logical and causal dependencies in the dataset.

\textit{2)} This makes it hard to preprocess data and instead encourages a ``first save everything and analyze it later'' approach, since it is difficult to anticipate what data and in which format it will be needed.
Simply first storing and then processing all available data comes with a high price and can easily lead to disproportionate computational efforts.
Let's just think of time series where huge sequences of data points needs to be stored and analyzed, like for ocean tides, stock values, and weather data. 
The quantity of data being generated from these sources makes time series storage, indexing ({\em e.g.,} signature files, B-trees), and processing highly challenging~\cite{DBLP:conf/seke/0001FNMKT14}~\cite{DBLP:conf/models/0001FNMKBT14}.
Considering this together with the previous point means that we have to restart the processing pipeline on a continuously growing dataset, although we might more and more process unused data.  
If data could be inferred (with a domain rule) from already computed knowledge, or if we knew based on expert knowledge that we don't need this data, there would be no need to store it.

\textit{3)} The data-centric approach that lies at the heart of todays analytics makes it difficult to anticipate the effects of actions.
Take load forecasting for electrical cables as an example. 
Classical analytics can tell us about historical load patterns and perhaps be used to predict the usual load of this cable during a specific time of the year. 
However, such techniques cannot help us to anticipate what happens if, lets say, we have to disconnect a related cable for maintenance reasons.
Such complex what-if questions need models developed by power grid experts describing the underlying electrical laws, grid topology, and so on.
Moreover, to simulate what-if questions todays pipeline-based approaches need to recompute the whole output dataset based on the changed input since there is no semantic description of causes and effects. 

\textit{4)} The knowledge of how to turn raw data into valuable insights is scattered around different places, {\em e.g.,} different map-reduce functions, and is therefore difficult to extract and maintain.
In other words, the link between raw data and knowledge is hidden in the implementation. 

\section{Requirements for Live Analytics of Complex Domains}
\textbf{All of the examples of~\ref{sec:bigdata_pathology:examples} have in common that they require to make decisions quickly ---often based on very small changes in the dataset--- and that making these decisions heavily rely on the combination of different analytic tools, data, ML, and, most importantly, domain knowledge to correctly assess the impacts of decisions.}
Most of todays analytics ---even if extended with stream processing capabilities~\cite{DBLP:conf/models/0001MFNKT15}--- are not well suited for operations which require to change only small parts of the dataset and need to combine various different analytic tools on these small changes.
Sequential decision-making in such complex domains with a high degree of uncertainty is inefficient. 
For example, not every information in a dataset might be useful for an analytics task.  
Some data might be just completely irrelevant for certain tasks, others might be even misleading, {\em e.g.,} peak values or measurement errors. 
Knowing what data is important and in which format, requires deep knowledge and expertise of the underlying domain.  

It requires to know the causes and effects of a domain (what-if models) to convert raw data into actionable intelligence or valuable knowledge. 
Besides the challenges related to the amount of data, the particular application domain knowledge, {\em e.g.,} formulated as scientific models, must be combined with raw data in order to create a sustainable refinement process.
Unlike the refinement process for oil, processing data of complex domains isn't sequential.
The results of one processing step ---gained knowledge--- usually influences earlier steps, so that a feedback loop (similar to reinforcement learning) is necessary to understand and interpret some of the results and to efficiently reuse already extracted knowledge. 
\textbf{The refinement process of data is a continuous process rather than sequentially storing everything and then processing it.}
To realize such continuous refinement process, data must be connected with domain knowledge and learning rules.
Defining causal relationships within data, based on a deep understanding of a domain, allows to anticipate consequences and causes of actions and to learn new causal relationships hidden in data.
\textbf{Such description, expressed in form of a domain model, can be seen as a view on raw data and can drive analytics that continuously refines data, always reflecting the latest stage of the domain knowledge.}
This is depicted in Figure~\ref{fig:todays_analytics} b). 

\section{Models to Connect Data, Domain Knowledge, and Learning}
An important characteristic of the nature of knowledge is the representation of relationships between things and finding these relationships is part of learning.
Therefore, techniques to organize, store, and process data according to a structure that concretizes these relationships are essential.

\subsection{Structuring Knowledge}
\label{sec:structuring_knowledge}
Over time different languages and formalisms to represent data and knowledge have been developed for different purposes.
For instance, relational databases are built since decades on top of relational algebra that embodies relationships by means of join operators. 
Graphs are used as a natural way to organize knowledge in form of nodes and edges as computable structures.
Ontologies are another example of a formalism to define types, properties, and relationships between entities of particular domains. 
These approaches have in common that they structure data by defining domain concepts and relationships between them in some way or another. 
This is equally important for both analytics and ML tasks. 
However, without a shared structure both domains can't cross-fertilize each other.

Such shared structure can be defined with models in the sense of model-driven engineering, which are widely used to formalize and structure the knowledge of complex domains.
Models are abstractions of a domain representing its essential rules, concepts, attributes, and relations between them. 
Such models are an efficient way to tackle high domain complexity by defining the domain knowledge and structure on a central place, instead of spreading it over the implementation of many functions. 

\subsection{How Models can drive Data Analytics}
\textbf{The rational behind model-driven analytics is to, based on a domain model, continuously refine raw data into a knowledge base, by connecting domain knowledge, data, and learning rules.}
A deep understanding of data requires the knowledge of domain experts.
We cannot just guess impacts of complex actions, neither can we predict the future based on the past if something never happened before.
But, if we know ---even in a simplified way--- the internal logic (causality, semantics) of the domain, we can combine past information, current observations, and domain knowledge to predict what will happen if we take a certain action or if a certain event would occur.
If we come back to the electricity grid example, we can hardly learn electrical laws and phenomenas by just observing raw data.
However, a model can describe knowledge which experts explored and collected over a long period.  
Analytic processes leveraging such models would be able to predict impacts of actions and events. 
\textbf{Moreover, the modelled causal relationships within data allows to refine only the necessary parts of data instead of sequentially recalculating everything (drawback 1).} 
For example, disconnecting an electricity cable only affects a couple of other cables and connected households.
This could be inferred from a topology model and electrical laws. 

The model forms a domain view of data relevant for specific analytics.
Instead of blindly storing and processing everything, intelligently filtering and preprocessing raw data based on domain expertise and anticipation of what we want to do with this data, can significantly reduce what we need to store and what we need to process.
\textbf{This avoids the ``store everything and analyze it later'' strategy promoted by most of todays pipeline-based analytics (drawback 2).}  
Moreover, it defines an unified view of heterogenous data, coming potentially from various different  sources, and its semantics. 

Beyond the sheer amount of data the complexity in such systems comes to a great extend from complex domain notions, the dependencies between them, and various rules to extrapolate meaning out of various measurements.   
\textbf{Experts can describe their knowledge of actions and effects in form of models, which then can be used in analytics.
This lays the ground for what-if analysis (drawback 3).}

\textbf{Moreover, these rules are defined in a central place, together with the domain structure of data, instead of being spread over the implementation of the analytic tasks (drawback 4).}

To sum up, models can be used, based on the domain knowledge of experts, to explicitly define the semantic of raw data, {\em e.g.,} in form of domain formula, mathematical models, and learning rules.
In addition, the semantic of data can tell us what data we need for the analytics, which information we might be able to infer from already stored data, and what data can be ignored. 
This allows to only store what is actually needed and also to know what knowledge (already processed data) need to be updated. 

\subsection{Model-Driven Analytics: A Continuous Refinement Process}
We call a model containing the essential concepts of a domain and the relationships between them, \textit{``knowledge graph''} (KG).
This graph represents the understanding of a domain, which is locked up in domain experts' heads and is hidden in raw data.
The model is like a \textit{view} on raw data that leverages various filtering and processing steps to map raw data to conceptual knowledge.
These abstractions are built to bridge the gap between a conceptual view and raw data.

To elicit such domain understanding out of data, different analytic tools and algorithms need to be combined and enriched with knowledge of domain experts. 
This is somehow similar to the idea of deep learning where different algorithms are used to model high-level abstractions in data, leveraging multiple processing layers with complex structures~\cite{Bengio:2009:LDA:1658423.1658424}. 
Following the nomenclature of \textit{knowledge graph}, we call the structure which connects various analytic tools, algorithms, and learning rules with the persistent raw data \textit{``persistent backend graph''} (PBG). 

\begin{figure*}
	\centering			
	\includegraphics[scale=0.45]{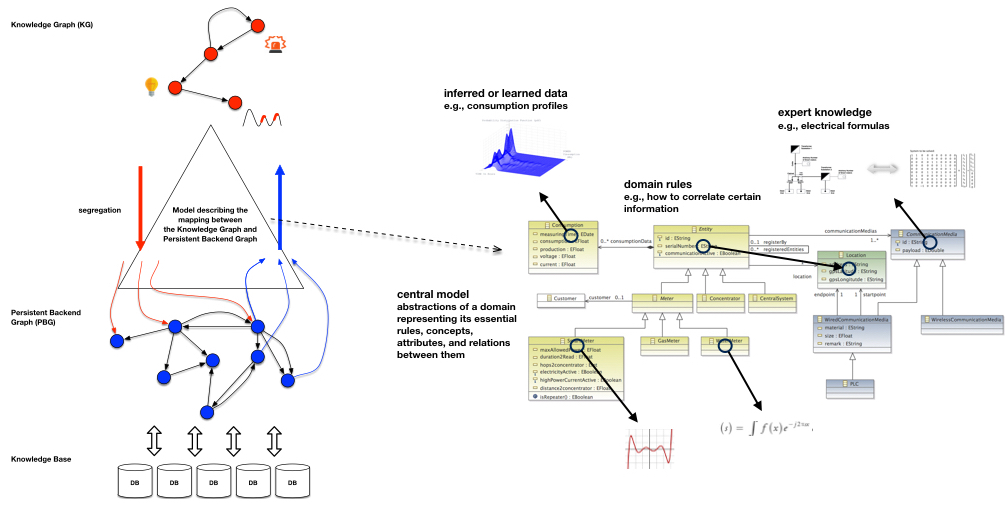}
	\caption{Models as the heart of knowledge refineries: models can describe the mapping of raw data into knowledge}
	\label{fig:models_as_refineries}
\end{figure*}

These two graphs are logically connected in the sense that the KG is the result of a refinement process of the information contained in the PBG and the PBG is the source for the KG. 
Viewed from top to down an element in the KG can be composed of various PBG elements.
The PBG elements are interconnected to be able to solve complex learning and analytic tasks. 
Elements of the PBG are usually persisted in form of raw data and additional information, such as meta data, used for analytics and learning processes.
Viewed from bottom to up several elements from the PBG are combined and aggregated to reflect a part of the domain knowledge.
As depicted in Figure~\ref{fig:models_as_refineries} the refinement process of model-driven analytics synchronizes these two views based on a model. 
Unlike it is the case for sequential processes, knowledge (or changes) injected via a modification of the KG is transparently mapped to the PBG, which immediately reflects this modification and makes it possible to reuse the gained knowledge. 
This transparent and continuous synchronization of the PBG and KG is one of the major benefits of model-driven analytics. 
It ensures a view (the KG) containing all the available ``knowledge'' of the domain.
\textbf{Model-driven analytics defines a transparent and continuous process of decomposing knowledge into various analytic elements which can help to decompose the complexity of understanding as a composition of various analytic tools.}

\section{Enabling Model-Driven Analytics}
In the following we discuss a number of technologies, which we consider as particularly important to enable model-driven analytics. 

\textit{Modeling languages}: To describe a mapping between raw data and domain knowledge it is necessary to extend the expressiveness of todays modelling languages. 
Besides describing the structure of data we also need a way to describe the knowledge of domain experts.
This includes statistical metrics, extrapolation methods, complex physical phenomenas, as well as ML techniques.
Listing~\ref{lst:profiler} shows as an example the definition of a consumption profiler as we use it in  a smart grid monitoring project.
\begin{lstlisting}[label={lst:profiler}, caption={Definition of an electric consumption value profiler},basicstyle=\ttfamily\small]
class ConsumptionProfiler {
    with "GaussianMixture" 
    with resolution "1week"
  dependency consumption: Consumption
  input "consumption | =energyConsumed"
  input "consumption | =HOURS(timestamp)"
  output probability: Double }
\end{lstlisting}
The example shows a \texttt{ConsumptionProfiler} class.
The clause \texttt{with "GaussianMixture" with resolution "1week"} specifies that the profiler uses a gaussian mixture algorithm and builds weekly profiles. 
The profiler class declares a dependency to a class \texttt{Consumption}, where it uses the value \texttt{energyConsumed} for building the profiles.
\texttt{HOURS(x)} specifies an hourly resolution of the gaussian mixture. 
The output property yields the probability of a given value to be ``correct'' according to the underlying gaussian mixture.
This shows how learning algorithms can be defined alongside domain models, allowing experts to specify where they expect to gain useful insights from.
 
\textit{Big data technologies}: Performance and storage efficiency are critical points for model-driven analytics. 
These are not independent but mutually influence each other.
The more ``unnecessary'' information we store the more data we have to analyze and the less performant this will be.
Instead of blindly storing and processing everything, intelligently filtering and preprocessing raw data based on domain expertise and anticipation of what we want to do with this data can significantly reduce both processing and storage costs.

\textit{Temporal data}: Most data is inherently temporal: 
from our smart grid example, through self-driving cars, financial applications, medical systems, to insurance applications. 
Modeling, storing, and analyzing temporal data is challenging.
Visible attempts in this direction are temporal databases to efficiently store and query time series, which are especially important for many analytic tasks. 
An efficient handling of time from modeling, over storage, to querying temporal data is therefore an important aspect of analytics and can significantly speed-up time series analysis and forecasting.

\textit{Machine learning}: Effective analytics has to allow us to learn from the data we collect.
Examples are classification, clustering, density estimation, anomaly detection, and hypothesis evaluation. 
ML is a powerful tool for prediction, which in turn is key for making sustainable decisions.  
ML expressed on top of models allows domain experts to directly express which correlations they want to learn and thereby adds an additional layer of semantic to raw data.

\textit{What-if analytics}: The exploration of what might happen if this or that action would be taken is fundamental for decision-making and \textit{prescriptive analytics}. 
This goes beyond statistical forecasting and is difficult to achieve by focusing only on current and historical data, which makes it hard to anticipate ---or simulate--- effects.
This also counts for ML, which can only find patterns and relationships which are already present in current or historical data. 
For example, ML techniques can hardly help if we want to know what impacts it has if a certain cable in the grid would be damaged.
However, considering, in addition to data, a model capturing the understanding of actions and effects of a given domain, can allow to simulate such scenarios. 

\section{Where we are: A Model-Driven Analytics Case Study}
\begin{figure*}	
	\centering			
	\includegraphics[scale=0.40]{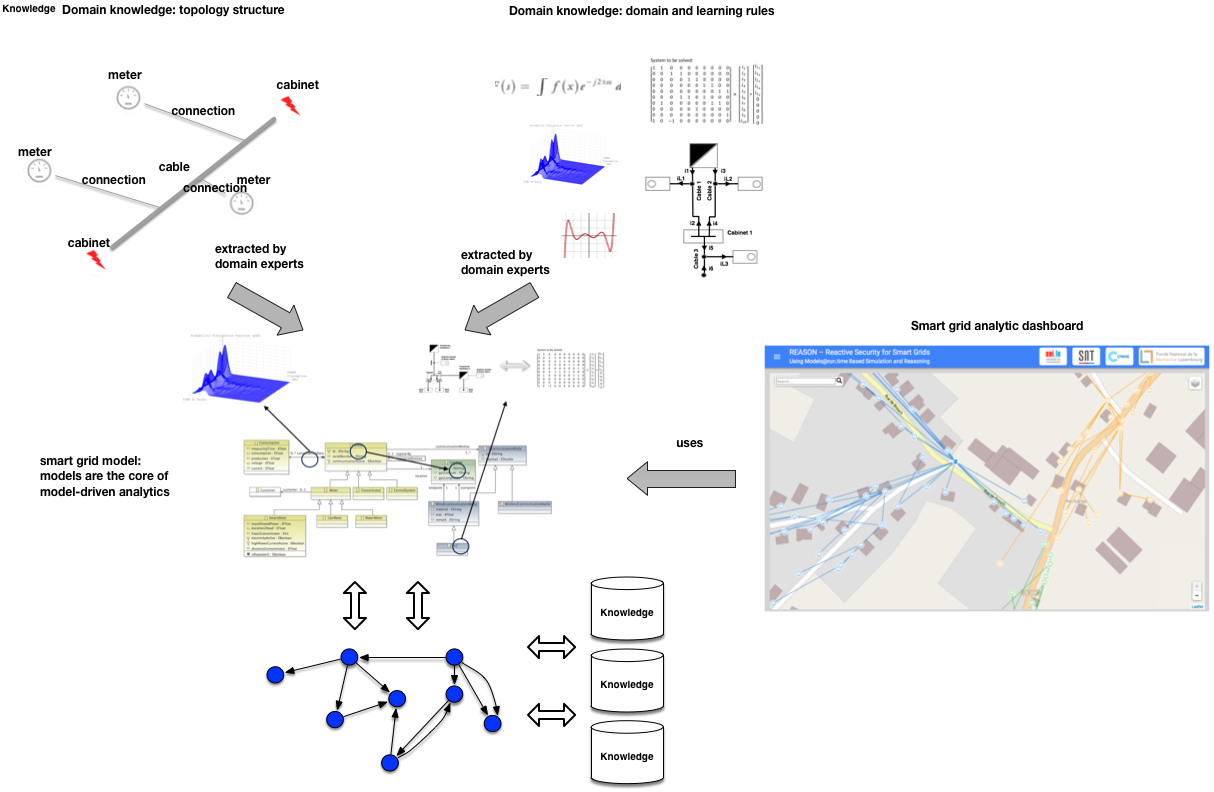}
	\caption{Model-driven analytics case study: smart grids}
	\label{fig:mda_case_study}
\end{figure*}

In this chapter we discuss a real-world case study, a smart grid monitoring system, where we apply and develop model-driven analytics together with our industrial partner Creos Luxembourg.
Considering the criticality of the electricity grid for modern societies, monitoring smart grid infrastructures become increasingly important to ensure their reliability and safety. 
This requires advanced data analytics. 
Together with experts from Creos we extracted the schematic structure of the physical grid with its cables, smart meters, concentrators, transformers, etc., into a model~\cite{DBLP:conf/smartgridcomm/0001FKTPTR14}.
Then, we augmented this model with additional domain knowledge, such as mathematical formulas, extrapolation, inference and learning rules.
The corresponding scheme is depicted in Figure~\ref{fig:mda_case_study}.
This augmented model is at the heart of our smart grid data analytics. 

A first challenge we were facing is the large amount of sensor data measured by smart grids, {\em e.g.,} customers' electricity consumption values, temperature, status reports, and electrical loading.
Following the idea to only store what is needed for the analytics (see~\ref{sec:structuring_knowledge}) to not get caught in the big data trap, we don't blindly save all of this data.
Instead, we approximate these values with mathematical polynomials and only store the polynomials.
This preprocessing becomes possible because domain experts know the semantic of data and can assess if a polynomial representation, which comes with a certain error rate, is appropriate for certain values or not.  
Besides significantly decreasing the required storage space, analysing a polynomial instead of thousands or millions of values severely speeds-up analytics. 
This addresses drawback 2 of~\ref{sec:drawbacks}. 
We showed a possible compression rate of 99\% for constant values, a range between 73\% to 46\% for IoT datasets, and 33\% to 10\% for random signals, based on different datasets: constant, electricity, temperature, luminosity, music files, random.
Random read operations, which are critical for analytics, could be improved in average by a factor of 40 to 60~\cite{7338239}. 
Furthermore, by considering time as a first-class entity together with a novel concept to model and process time series we could significantly improve a load forecasting case study with Creos.

To address drawback 1 of~\ref{sec:drawbacks} we are working on allowing domain experts to seamlessly define the semantic of data. 
This makes it possible to only update the knowledge which is affected instead of recalculating everything. 
One example is the automatic detection of suspicious consumption values.
Given the big amount of meter data this is rather challenging, especially because it heavily depends on the context (working days, weekends, temperature, time).
We developed a live ML approach which continuously learns context-dependent consumption profiles, classifies them, and selects the most appropriate one according to the context.
Whenever a new value arrives only the affected profiles are incrementally updated. 
Experiments with real data showed an accuracy between 83\% to 93\%~\cite{hartmann2015suspicious}.
Detecting problems in the grid is one side of the coin. 
The other one is to find appropriate reactions.
This makes it necessary to simulate different actions and to decide which is the most appropriate.
We are working on efficient data structures to independently simulate different actions and on extended modeling languages to define different actions and the causes and effects of these.
This addresses drawback 3 of~\ref{sec:drawbacks}. 

\section{Related Work}
Model-driven analytics pursues the idea of model-driven engineering further and brings it to another domain: data analytics.
Furthermore, it can be seen as an advancement of the models@run.time~\cite{morin2009models} paradigm which promotes the usage of runtime models to reason about the state of a running system. 
Similar to this paradigm, model-driven analytics suggests to use domain models as an abstraction which is simpler than the reality. 
Whereas models@run.time abstracts the state of complex cyber-physical systems, model-driven analytics abstracts the expert knowledge of a domain in form of domain laws, mathematical formula, and learning rules to bring deep understanding to raw data.
Bishop~\cite{bishop2013model} presents a model-based ML approach together with a modelling language to express ML problems on a higher level.
This goes in the same direction than what we intend to achieve with model-driven analytics.
Here too, models are used as higher-level abstractions to drive complex processes.
Crankshaw {\em et al.,}~\cite{crankshaw2014missing} also identified the lack of semantic descriptions of input and output in typical analytics pipelines as a problem.
With Velox, they propose a solution to model serving and management for the Spark~\cite{zaharia2012resilient} stack. 
Velox manages the ML lifecycle from training on raw data to predictions that the models inform. 
Like we suggest with model-driven analytics, they leverage models to bring more semantic to the level of data and analytics. 

Model-driven analytics combines various areas of research, such as software engineering, ML, databases, big data, modelling, and analytics.
The concept of model-driven analytics is yet in an early stage and opens many interesting research challenges.
To name just a few: advanced languages could allow domain experts to specify their knowledge in  comfortable ways, {\em e.g.,} in natural language. 
Another challenge is the integration of advanced ML techniques, like reinforcement learning, to improve analytics. 
Closely linked to this are natural interfaces enabling domain experts to interact with an analytics system, {\em e.g.,} to evaluate results of the system for feedback loops~\cite{DBLP:conf/modelsward/Moawad0FNKB15}.

\section{Conclusion}
With model-driven analytics we promote the idea to use extended domain models to define the semantic of data in form of domain laws, mathematical formula, and learning rules to gain new insights from raw data. 
It brings domain knowledge in form of models to the level of data. 
These models are building a bridge between domain understanding and the actual raw data representing the sensed reality.
Like refineries transform crude oil into an essential energy source, models are defining the necessary semantics to transform raw data into valuable or actionable insights.

\section*{Acknowledgment}

\small
The research leading to this publication is supported by the National Research Fund Luxembourg (grant 6816126) and Creos Luxembourg S.A. under the SnT-Creos partnership program.
\normalsize

\bibliographystyle{IEEEtran}
\bibliography{IEEEabrv,computerbib}

\end{document}